\pdfoutput=1
\documentclass[fleqn,usenatbib]{mnras}

\usepackage{newtxtext,newtxmath}
\usepackage[T1]{fontenc}

\DeclareRobustCommand{\VAN}[3]{#2}
\let\VANthebibliography\thebibliography
\def\thebibliography{\DeclareRobustCommand{\VAN}[3]{##3}\VANthebibliography}

\usepackage{graphicx}	% Including figure files
\usepackage{amsmath}	% Advanced maths commands

\usepackage{amssymb}	% Extra maths symbols
\usepackage{threeparttablex} % for "ThreePartTable" environment
\usepackage{xcolor}
\usepackage{diagbox}
\title[The orbital evolution of Atira asteroids]{The orbital evolution of Atira asteroids}

\author[H. T. Lai et al.]{
H. T. Lai,$^{1}$\thanks{E-mail: wingip@astro.ncu.edu.tw}
 W. H. Ip$^{1,2}$
\\
$^{1}$Institute of Astronomy, National Central University, Taiwan\\
$^{2}$Department of Space science and Engineering, National Central University, Taiwan\\
}

\date{Accepted 2022 October 15. Received 2022 September 20; in original form 2021 October 26}

\pubyear{2022}

\begin{document}
\label{firstpage}
\pagerange{\pageref{firstpage}--\pageref{lastpage}}
\maketitle

\begin{abstract}
Asteroids having perihelion distance $q$ $<$ 1.3 AU are classified as near-Earth objects (NEOs), which are divided into different sub-groups: Vatira-class, Atira-class, Aten-class, Apollo-class, and Amor-class. 2020 $AV_2$, the first Vatira (Orbiting totally inside Venus' orbit) was discovered by the Twilight project of the Zwicky Transient Facility (ZTF) on January 4, 2020. Upon the discovery of 2020 $AV_2$, a couple of orbital studies of the short-term orbital evolution of 2020 $AV_2$ have been performed and published \citep[e.g.][]{Marcos2020, Greenstreet_2020}. In this present work, we performed an assessment of the long-term orbital evolution of known near-Earth objects and known Atiras under the Yarkovsky effect by using the \textit{Mercury6} N-body code. We considered not only planetary gravitational perturbation but also the non-gravitational Yarkovsky effect. Our calculation shows that the NEOs have generally two dynamical populations, one short-lived and the other long-lived. From our calculation, the transfer probabilities of Atira-class asteroids to Vatira-class asteroids for the first transition are $\sim$13.1 $\pm$ 0.400, $\sim$13.05 $\pm$  0.005, and $\sim$ 13.25 $\pm$ 0.450 $\%$ for different values of the Yarkovsky force (i.e. obliquity of 0, 90, and 180 deg.), respectively. It suggests that the radiation force may play some role in the long-term evolution of this asteroid population. Finally, our statistical study implicates that there should be 8.14 $\pm$ 0.133 Atira-class asteroids and 1.05 $\pm$ 0.075 Vatira-asteroids of the S-type taxonomy.
\end{abstract}


%%%%%%%%%%%%%%%%%%%%%%%%%%%%%%%%%%%%%%%%%%%%%%%%%%

%%%%%%%%%%%%%%%%% BODY OF PAPER %%%%%%%%%%%%%%%%%%

\section{Introduction}
\label{Chap:Chap1} 
Asteroids with perihelion distance $q$ $<$ 1.3 AU, aphelion distance $Q$ $>$ 0.718 AU \citep{bottke2002debiased} and semi-major axis $a$ $<$ 4.2 AU\footnote[1]{There is no exact outer boundary of the semi-major axis for NEOs, so we set semi-major axis within 4.2 AU, which is referred from \citet{bottke2002debiased}, \citet{greenstreet2012orbital}, and \citet{granvik2018debiased}.  } are classified as near-Earth objects (NEOs). And they are divided into different sub-groups by the following criteria: Amor-class (1.017 AU $<$ $q$ $<$ 1.3 AU), Apollo-class ($a$ $>$ 1.0 AU, $q$ $<$ 1.017 AU), Aten-class ($a$ $<$ 1.0 AU, $Q$ $>$ 0.983 AU), Atira-class (0.718 $<$ $Q$ $<$ 0.983 AU) that orbits totally inside Earth’s orbit, where $Q$ is the orbital aphelion distance. From numerical simulations, \citet{greenstreet2012orbital} proposed the presence of the Vatira-class asteroids, also called inner-Venus objects (IVOs) - with Q between 0.307 AU and 0.718 AU before the discovery of the first Vatira asteroid, 2020 $AV_2$, with \textit{H} $\sim$ 16.4 $\pm$ 0.77 \citep{bolin2020mpec,ip2022discovery}.

The largest source region to supply near-Earth asteroids to the inner solar system is the main asteroid belt located between Mars and Jupiter. Some strong resonances can push asteroids inward or outward by pumping up their eccentricity value to reach planet-crossing orbit. They include the $\nu_6$ secular resonance at 2.2 AU, which is the inner boundary of the Main Belt \citep{morbidelli1994delivery}, as well as the 3:1 mean-motion resonance with Jupiter at 2.5 AU \citep{wisdom1983chaotic}. Another possible source is the fragments in the asteroid families generated after large catastrophic collisions. During the long-term orbital evolution, small asteroids with diameter $D$ $<$ 40 km subjecting to tiny thrust from thermal radiation (e.g. the Yarkovsky effect) could migrate into a strong mean-motion region leading to escape from the main asteroid belt \citep{bottke2006yarkovsky}. Also, the Jupiter-family comets which orbital periods are smaller than $P$ $<$ 20 years, are a source of near-Earth objects due to resonance perturbation, that was investigated by \citet{levison1997kuiper}, and \citet{bottke2002debiased}.

 In a previous comprehensive numerical simulation, \citet{granvik2018debiased} identified the Jupiter-family comets and six source regions of NEOs from the main asteroid belt: the  $\nu_6$, 3:1J, 5:2J, 2:1J mean-motion resonance with Jupiter, Hungaria and Phocaea asteroid populations by using calibrated data of NEOs detected by Catalina Sky Survey (CSS) during 2005-2012. It was shown how a steady state orbital distribution can be established in which main belt asteroids can be transported to the inner solar system due to the influence of the mean motion resonance and planetary close encounters. While the median time of NEOs from the main asteroid belt to reach Earth-crossing orbit takes about $\sim$ 0.5 Myr and $\sim$ 1.0 Myr for $\nu_6$ secular resonance and 3:1 resonance with Jupiter, respectively \citep{morbidelli2002origin}, \citet{gladman1997} indicated that once asteroids were delivered to the terrestrial planet-crossing region, thier dynamical lifetime will be determined by frequent planetary close encounters, impacts with the Sun and planets or ejection onto escape orbit by the planets.

Upon the discovery of 2020 $AV_2$, its short-term orbital evolution have been investigated by \citet{Marcos2020} and \citet{Greenstreet_2020}. They indicated that 2020 $AV_2$ will be possibly trapped by the 3:2 resonance with Venus and the influence of the Zeipel–Lidov–Kozai oscillation \citep{von1910application, lidov1962evolution,kozai1962secular} will prevent it from close encounter with Mercury and Venus. In addition, 2020 $AV_2$ probably has switched from the Atira-class to the Vatira-class in the past $\sim$ $10^5$ years. For the orbital dynamics of Atira-class asteroids, \citet{ribeiro2016dynamical} studied the dynamical evolution of the inner Earth region to identify the stable region with eccentricity \textit{e} $<$ 0.2 and inclination \textit{i} $<$ $30^o$ between Mercury and Venus (semi-major axis of 0.5 AU $\sim$ 0.65 AU), which overlaps the orbital region of Vatira-class asteroids and 2020 $AV_2$.

 Model calculation of NEOs with absolute magnitude of 17 $<$ $H$ $<$ 25 from \citet{granvik2018debiased} predicts the fraction for Vatira-class to be $0.25_{-0.01}^{+0.01}$ \% when we consider 2020 $AV_2$. According to the NEOs models described in \citet{granvik2018debiased}, the likely possible origins of 2020 $AV_2$ are $\sim$ 77\% probability from $\nu_6$ secular resonance, $\sim$ 18\% probability from 3:1J resonance, and $\sim$ 4\% probability from Hungaria asteroid population.
 
For long-term orbital evolution, the Yarkovsky effect as a non-gravitational force can affect the direction of asteroid drifting owing to the thermal lag between the absorption of the solar radiation in the morning part and the thermal emission arising on an asteroid in the evening part. And the diurnal variation of the surface temperature distribution on the asteroid surface can thus play an important role in the delivery of meteorites and the dynamics of small bodies in the Solar system \citep{farinella1998meteorite}. The physical properties such as density, albedo, obliquity, thermal inertia, and diameter of asteroids have to be taken into consideration in the Yarkovsky effect. Here, since the seasonal component of the Yarkovsky effect is much smaller than the diurnal component, we assume the orbit-averaged drift rate as being dominated by the diurnal effect \citep{rubincam1995asteroid,farinella1998meteorite,vokrouhlicky1999yarkovsky}, only when the rotation is very slow, or the spinning obliquity is almost 90 degrees, would the seasonal effect play the dominant role. However, compared to the diurnal effect, the displacement of the semi-major axis induced by the seasonal effect is much smaller for an regolith-covered object with the obliquity of 90 degrees at the initial semi-major axis of 2.5 AU in 10 Myr according to \citet{xu2020asteroid}. Therefore, the seasonal Yarkovsky effect is not included for the obliquity of 90 degrees. In other words, an asteroid with the obliquity of 90 degrees is regarded as being without Yarkovsky force in our simulation.

The paper is organized as follows. In section \ref{Chap:Chap2}, we describe the numerical procedure for investigating the dynamical evolution of known near-earth objects and Atira-class asteroids. And we present the results from the simulations in section \ref{Chap:Chap3}. Finally, we give a summary and assess the implication of our study.
 
\section{Numerical procedure}
\label{Chap:Chap2}
In order to study the dynamical evolution of near-earth objects, we performed two numerical simulations by using the \textit{Mercury6} N-body code with hybrid symplectic integrator \citep{Chambers1999}. Because the non-gravitational Yarkovsky effect is included in the software package as a subroutine, which is described in \citet{zhou2019orbital}. All test particles in the simulation are influenced by not only the gravitational force of the eight planets (from Mercury to Neptune) and the Sun but also by the non-gravitational Yarkovsky effect. For the analysis of the result of different sets of simulations, asteroids having perihelion distance $q$ $<$ 1.3 AU and aphelion distance $Q$ $>$ 0.718 AU are classified as near-Earth objects (NEOs). However, as some test particles with NEOs' orbital properties may migrate outward beyond the asteroid main belt, we restricted the semi-major axis of the NEOs to be within 4.2 au, which is the same criterion used in \citet{bottke2002debiased}. As for the Vatira-class asteroids (0.307 AU $<$ $Q$ $<$ 0.718 AU), they are classified separately and not as Near-Earth objects since their orbits are confirmed within Venus' orbit.

\subsection{Run A: The dynamical evolution of Near-Earth objects}
\label{run_A}
 In the first set of simulations, we queried 25811 identified NEOs from JPL on March 30 in 2021. As for the selection criteria of orbital elements, we filtered the sample candidates by the following conditions: (1) The number of observations must be larger than 10 to ensure the orbital elements of the asteroid have been well-obsered. (2) The semi-major axis of the asteroid is within 4.2 au, as defined by \citet{bottke2002debiased}, \citet{greenstreet2012orbital}, and \citet{granvik2018debiased}. (3) The orbit of the asteroid is prograde, namely inclination $<$ 90 deg. In addition, we excluded the Vatira-class asteroid (see section \ref{run_B}). Therefore, the remaining number of effective NEO samples is 24851 with the absolute magnitude of 9.4 $<$ $H$ $<$ 33.2. According to the orbital properties of different classes of near-earth objects, we used Monte Carlo method to select 3020 known near-earth objects as samples (see Table \ref{table:Tab1.0}). Because the total known number of Atira-class asteroids was only 23 at the time, we selected 20 of them as the samples based on the reasons described in section \ref{run_B}. As for Aten-class, Apollo-class, and Amor-class asteroids, we selected 1000 as samples for each class because the samples numbers of individual classes are large enough to represent the known NEO populations. Furthermore, to ensure that our samples well-represent each NEO sub-class, we used the Kolmogorov-Smirnov test (hereafter KS test) to qualify the similarity between selected samples and known JPL population. Here, we set a significant value of 0.05 for KS test. If the $p$-value given by KS test is smaller than the significance value, and the two populations are statistically different. The KS test was applied to the orbital elements of the semi-major axis, eccentricity, inclination, and absolute magnitude between the selected samples and the corresponding JPL sub-groups, we obtained $p$-values, which are over 0.05, supporting the assessment that the samples are well-selected to represent the known population in JPL database (see the detailed result in Appendix \ref{Chap:Append_A}). Following the statistical model of \citet{granvik2018debiased} for the relative proportions of different types of NEOs irrespective of the absolute magnitude $H$, the following weighting factors are needed: Amors = 40.0\%, Apollos = 55.23 \%, Atens = 3.55 \%, and Atrias = 1.22 \%. In order to compare the dynamical evolution of Near-Earth objects, the non-gravitational Yarkovsky effect isn’t included in this set of integration, and we only consider the gravitational influence of eight planets and the Sun to integrate each sample forward integration in 40 Myr. The output time interval is recorded every 50,000 years by the heliocentric coordinate system, the time step is set to be 4 days.

\subsection{Run B: The orbital evolution of the Atira-class asteroids under the Yarkovsky effect}
\label{run_B}
As noted before, there are totally 23 known Atira-class asteroids including 2020 $AV_2$ identified by Jet Propulsion Laboratory (JPL) Small-Body Database on October 31 2020. We didn't consider 2020 $AV_2$ in the simulation because it is classified as Vatira-class asteroid by our criteria. Also, we only considered the effective sample with the number of observations N > 10 to avoid cloning test particles with large uncertainty. Two of the 23 known Atiras-class asteroids were thus removed from the effective sample list. Therefore, there are the remaining 20 effective samples with 16.3 $<$ \textit{H} $<$ 25.0. In this set of simulation, we generated 100 clone particles per known Atira-class asteroid by the covariance matrix of six elements (i.e., $a, e, I, \omega, \Omega, M$) from the JPL Small-Body Database website. All the arbitrary values of orbital parameters of clone particles followed a Gaussian distribution with 1-sigma uncertainty. Also, each known Atiras-class asteroid was given three kinds of obliquities (e.g. 0, 90, 180 degrees) as far as the Yarkovsky force is concerned. While asteroids with rotation obliquity of 90 degrees are regarded as not being affected by the non-gravitational effect. In the second set of simulation (Run B), we carried out numerical integration for the 100 clone sets of each of the 20 different Atiras for 40 Myr under the gravitational influence of the eight planets from Mercury to Neptune and the non-gravitational Yarkovsky effect. The time step is 4 days, and the output interval time is every 10,000 years (see Table \ref{table:Tab1.0}). 

To assess the strength of the Yarkovsky effect, the corresponding drift rate is adopted from \citet{spoto2015asteroid} and \citet{bolin2018size};

\begin{equation}
    \frac{da}{dt}(D,a,e,\theta) = \left(\frac{da}{dt}\right)_{0}  \frac{\sqrt{a_0}(1-e_{0}^2)}{\sqrt{a}(1-e^2)}\frac{D_0}{D}\frac{\cos{\theta}}{\cos{\theta_0}}\left(\frac{AU}{Myr}\right),
    \label{eq1}
\end{equation}

where $a_0$ = 2.37 AU, $e_0$ = 0.2, $D_0$ = 5 km, and $\theta_0$ = 0 deg. in \citet{bolin2018size}. $\theta$ is the rotation obliquity of an asteroid and $D$ is the diameter of an asteroid. The drift rate of $(da/dt)_0 \sim 3.4 \times 10^{-4}$ AU $Myr^{-1}$ is for a 1 km-sized asteroid at 1 AU. The mean semi-major axis drift $(da/dt)_{0}$ has been directly measured on (101955) Bennu by optical astrometry and radar measurements from 1999 to 2011 \citep{chesley2014orbit}. The Yarkovsky force only acts on the effective diameter of asteroids between 0.1 m and ~40 km, so the thermal force acting on a very large or very small objects is negligible, such as planets and dust \citep{bottke2006yarkovsky}. For a very large asteroid, the thermal force is too small to change the orbit of the object, while there is no temperature difference on the surface of a very small object to produce thrust from thermal re-radiation since the penetration depth of the thermal wave is larger than the size of a very small asteroid \citep{bottke2006yarkovsky}. In general, the Yarkovsky-driven drift rate for large objects (i.e. kilometer-sized objects) can be described as $(da/dt) \propto D^{-1}$ depending on the diameter of the objects. After that, according to the relationship between absolute magnitude, geometric albedo, and size of asteroids described in \citet{harris1997revision}, the size of the asteroid is converted from absolute magnitudes $H$ with the geometric albedos assumed to be $p_v$ = 0.15, which is the typical value for S-type asteroids \citep{masiero2011main, masiero2015asteroid}. The strength of the Yarkovsky force increases with decreasing distance to the Sun owing to receiving more solar radiation. From a consideration of the rate of change of the angular momentum because of the radial distance-dependence of the thermal radiation thrust, the Yarkovsky effect can be found to be proportional to $a^{-0.5}$. Accordingly, the average drift rate is $(da/dt)_0 \sim 9.85 \times 10^{-4}$ AU $Myr^{-1}$ for a half-kilometer-sized sample with the rotation obliquity of 0 degrees in the simulation.

% table 1.0 
\begin{table}
	\centering
	\caption{The initial parameters of the simulations.}
	   \label{table:Tab1.0} 
       \begin{tabular}{lll} \hline
        Simulation & Run A & Run B  \\ \hline
        Integration time (Myrs) & 40 & 40 \\
        Time step (days) & 4 & 4 \\
        Output interval time (years) & 50,000 & 10,000\\
        Number of the effective samples & 3020 & 20  \\
        Class of Asteroid & NEOs & Atira-class  \\
        Yarkovsky force & No & Yes  \\ \hline
        \end{tabular}
\end{table}

\section{Results}
\label{Chap:Chap3} 
\subsection{Dynamical half-lifetime}
To evaluate the dynamical half-lifetimes of the test particles, we restricted the boundary of the dynamical region ($q$ $<$ 1.3 AU, $Q$ $>$ 0.718 AU and $a$ $<$ 4.2 AU) for NEOs to check whether asteroids are “alive” over the orbital integration time span. In other words, asteroids are regarded  as "dead" when leaving the dynamical region. The time variations of the asteroidal numbers remaining in specific orbital regions can’t be fitted to a single exponential law. This is because some members have rather long dynamical lives as demonstrated in the long-term calculations of \citet{evans1999possible}. As a consequence, a two-component expression as given in Eq.(\ref{eq2}) can give a better fit. Thus, we estimated the short-term and long-term half-lifetime for each class of NEOs population shown in Table \ref{table:Tab2.0}. The result provides the half-lifetime for asteroids with short and long dynamical lives, respectively, when the asteroidal population is reduced to 50\% of its original size.

\begin{equation} \label{eq2} 
N/N_0 = \alpha e^{\lambda_1 t} +  (1-\alpha)e^{\lambda_2 t}, ~~~ \text{with}~~~ \lambda_1 = \frac{ln2}{T_{1,1/2}} ~~ \text{and} ~~ \lambda_2 = \frac{ln2}{T_{2,1/2}}
\end{equation}

Where $N$ is number of NEOs, $N_0$ is the initial number of test particles in the simulation, $\alpha$ is a fitting parameter, $\lambda_1$ and $\lambda_2$ are decay constants, and $T_{1,1/2}$ and  $T_{2,1/2}$ are the half-lifetimes of NEOs. The subscript of decay constant and half-lifetimes for "1" and "2" represent the short-lived and long-lived components, respectively. As for the uncertainty, they were given by $\sigma_1 = \frac{T_{1,1/2}}{\sqrt{\alpha \cdot N_0}} $ and $\sigma_2 = \frac{T_{2,1/2}}{\sqrt{(1-\alpha ) \cdot N_0}} $ , respectively, according to \citet{horner2004simulations}. In order to find the best fitting curve, we used the reduced chi-squared statistic by varying $\alpha$ value between 0 to 1 in Eq. (\ref{eq2}) and estimated the short-term and long-term half-lifetimes when the reduced chi-squared is the minimum (see Table \textcolor{blue}{3}).

\begin{figure}
    \centering 
    \includegraphics[width=0.7\columnwidth]{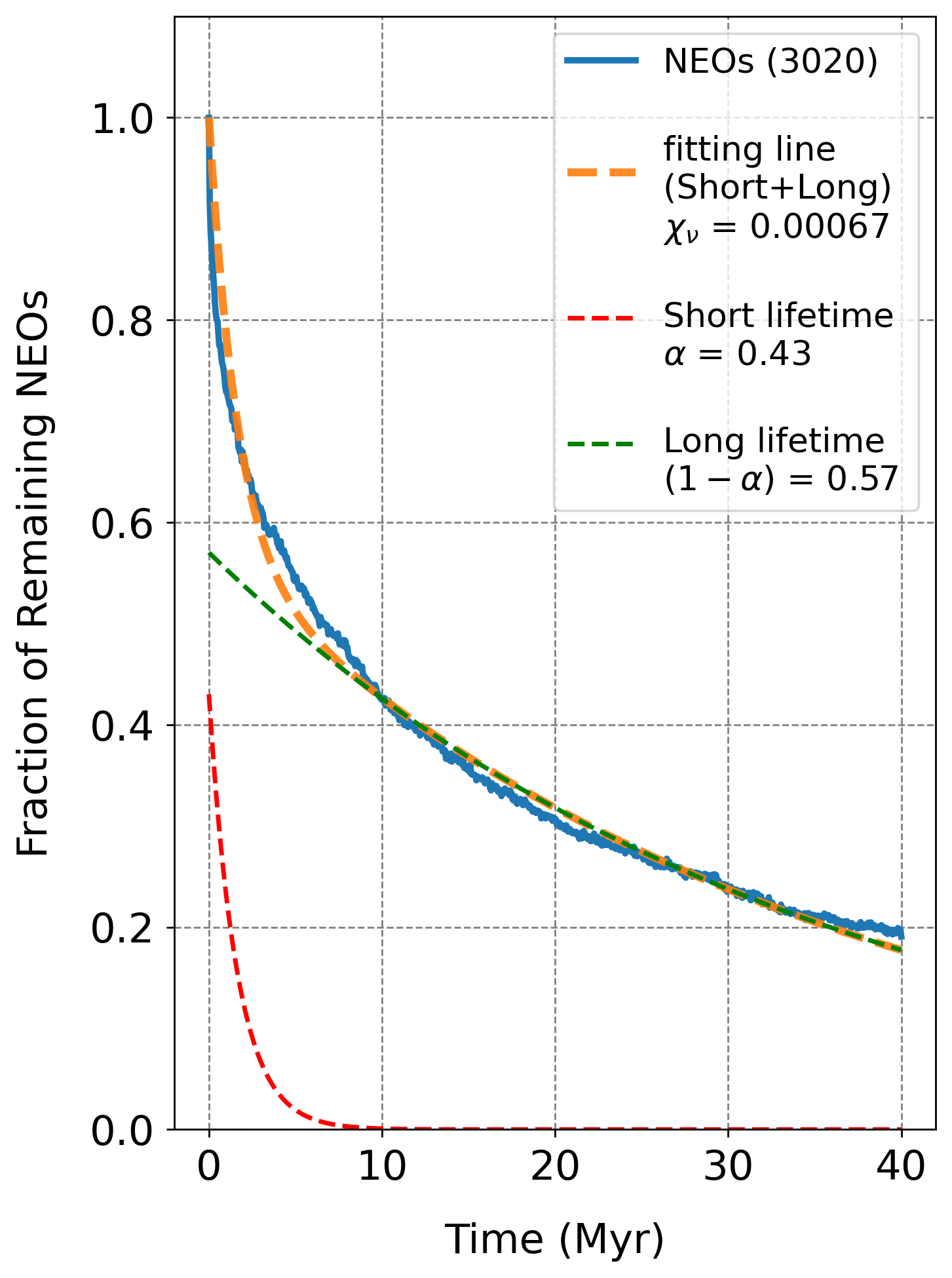}
    \caption{The dynamical half-lifetime of near-Earth objects in the Run A simulation, including Atira-class, Aten-class, Apollo-class, and Amor-class asteroids (see Table \ref{table:Tab2.0}). The orange dash-line is combined with short-term and long-term half lifetime, the green dash-line is the best fitting for the short-term half-lifetime, and the red dash-line is the best fitting for the long-term half-lifetime. $\chi_\nu$ is the reduced chi-square value calculated from the fitting line, and $\alpha$ is the fitting parameter calculated from Eq. (\ref{eq2}).}
    \label{fig1}
\end{figure}

\begin{figure}
    \centering 
    \includegraphics[width=1.0\columnwidth]{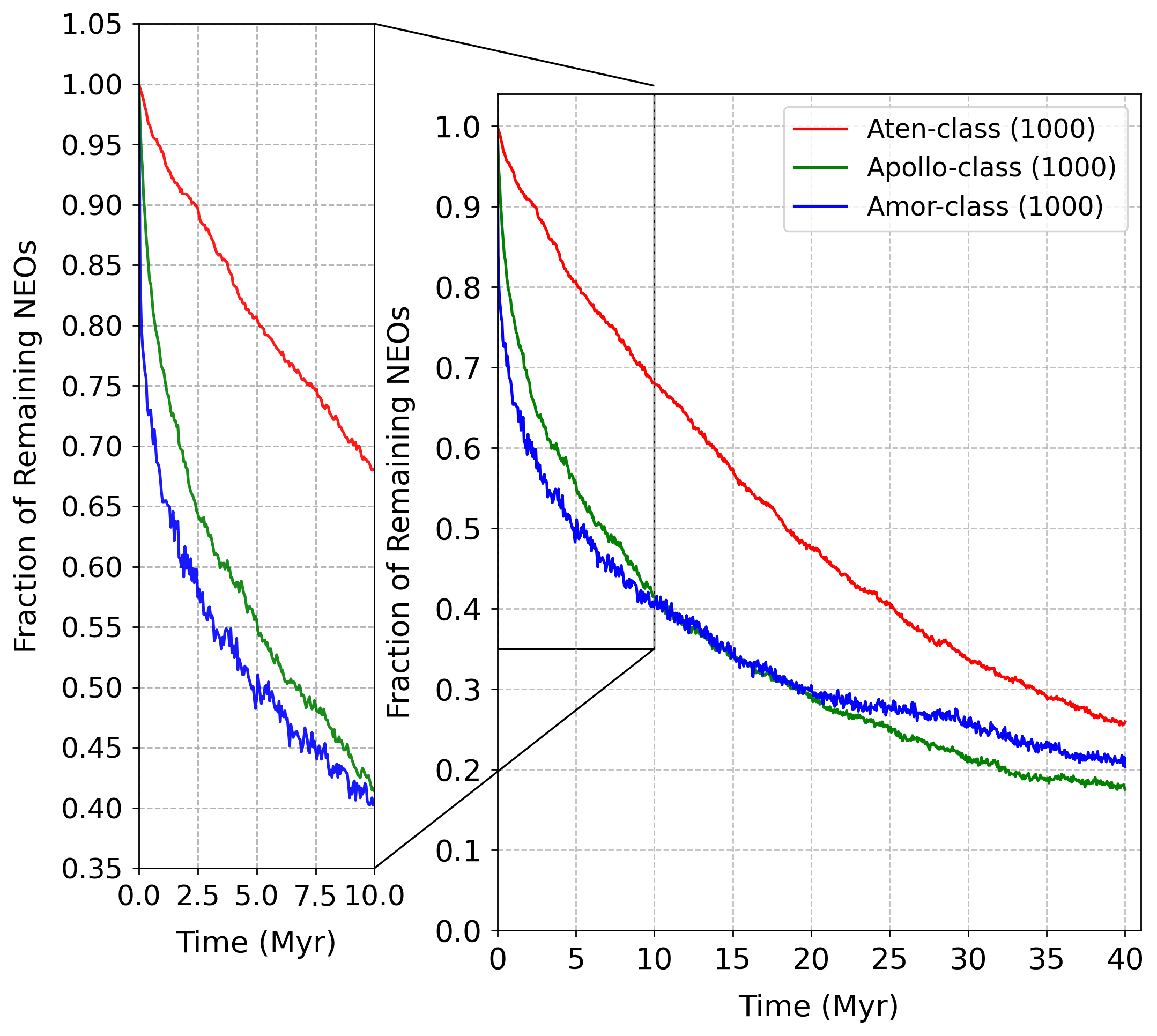}
    \caption{ The dynamical half-lifetime of the Aten-class (Red), the Apollo-class (Green), and the Amor-class asteroids (Blue) in the Run A simulation (see Table \ref{table:Tab2.0}).} 
    \label{fig2}
\end{figure}

Figure \ref{fig1} shows the time variation of the NEOs from which the curve can be fitted by a combination of a short-life population with a half-life of 1.12 Myr and a long-life population with a half-life of 23.74 Myr (see Table \ref{table:Tab2.0}). Note that one of the sub-groups of NEOs, namely the Amor-class, has much shorter short-term half-lifetime of $\sim$ 0.76 Myr than the other classes of NEOs (see Figure \ref{fig2}). This is because most Amor-class asteroids located at the boundary of the NEOs region can be easily ejected out in the short-term evolution. The Aten-class asteroids having a longer short-term half-lifetime than other classes of NEOs are located in the inner dynamical region, they need to spend more time migrating out of the dynamical region hence large lifetime. Furthermore, we found that the Amor-class asteroids have a longer long-term half-lifetime than the Aten- and Apollo-class asteroids. Similarity, the long-term half lifetime for the Atira-class asteroids with 90 deg obliquity (i.e., no Yarkovsky force) is very similar to that for the Amor-class asteroids. This effect is probably because the Amor-and Atira-class asteroids are not on Earth-crossing orbits and thus have fewer planetary close encounters than the Aten- and Apollo-class asteroids. And this result is in general agreement with the finding from \citet{gladman1997} that NEO dynamical lifetimes are influenced by frequent planetary close encounters.

\begin{figure} 
    \centering 
    \includegraphics[width=1.0\columnwidth]{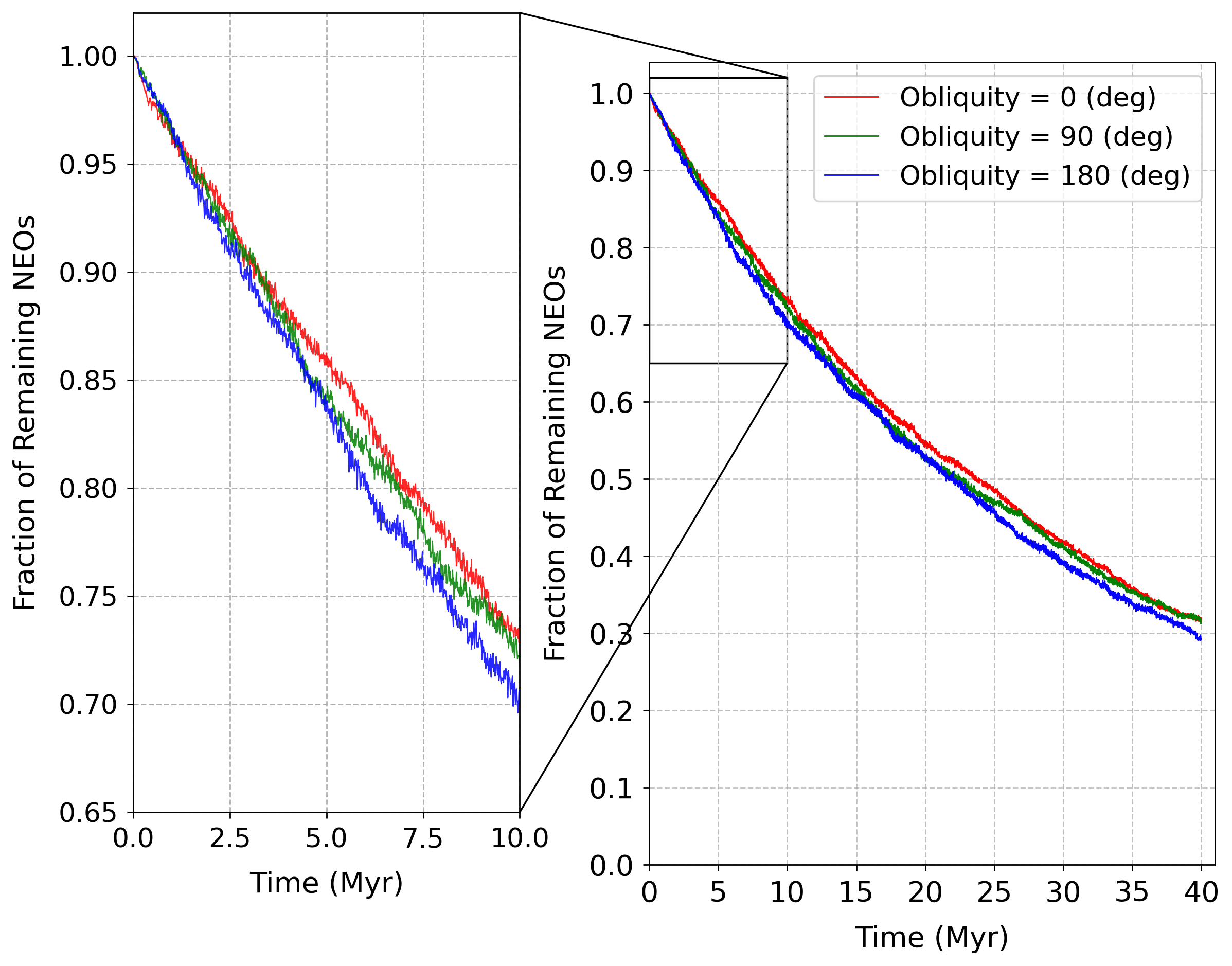}
    \caption{The dynamical half-lifetime of the Atira-class asteroid with the rotation obliquity of 0 (red), 90 (green), and 180 (blue) degrees was considered in the Run B simulation. (see Table \ref{table:Tab2.0}).  }
    \label{fig3}
\end{figure}

As for the Atira-class asteroids, Figure \ref{fig3} shows that the short-term half-lifetime for different values of the Yarkovsky force are $\sim$ 4.22, 8.15, and 3.61 Myr, respectively, and the long-term values $\sim$ 24.78, 28.26, and 24.11 Myr, respectively. We notice that the dynamical half-lifetime of the Atira-class asteroids without the Yarkovsky force is longer than the Atira-class asteroids with non-gravitational force of $ \dot a \sim \pm 9.85 \times 10^{-4} AU Myr^{-1}$, namely asteroids with the obliquity of 0 and 180 degree, since asteroids affected by non-gravitational force have more opportunities to migrate out of the dynamical region and reach planet-crossing orbit. Moreover, for a comparison of our calculated half-lifetimes with that of \citet{gladman1997}, the mean of the short- and long-term half-lifetimes are shown in Table \ref{table:Tab2.0}. The mean half-lifetimes of 12.4 Myr for NEOs in the Run A simulation is close to $\sim$ 10 Myr median NEO half-lifetime calculated by \citet{gladman1997}.
% The short-term half-lifetime \textbf{of $\sim$ 8.15 Myr} for Atira-class asteroids with the rotation obliquity of 90 degree is in agreement with the median half-lifetime of NEOs given in \citet{gladman1997}. 

% fitting result

% table 2.0 : half-lifeitme of NEOs
\begin{table*}
  \centering
  \footnotesize
  \caption[Summary of dynamical half-lifetimes of NEOs and the Atira-class asteroids with different values of the Yarkovsky force in the NEOs region.]{Summary of dynamical half-lifetimes of NEOs and the Atira-class asteroids with different values of the Yarkovsky force in the NEOs region (0.718 AU $<$ $Q$, $q$ $<$ 1.3 AU, and $a$ $<$ 4.2 AU). The forward integration time is 40 Myr.}
  \begin{threeparttable} %Comment section out if you don't want table notes
\begin{tabular}{cccccccccc} 
\hline
Class & Simulation  & Number & \textit{$\alpha$} $^a$ & \textit{$\chi_\nu$} $^b$  & \textit{$T_1$} $^c$ ~(Myr) & \textit{$\sigma_1$} $^d$~(Myr) & \textit{$T_2$}  $^e$~(Myr) & \textit{$\sigma_2$} $^f$~(Myr) & $\bar T$ $^g$~(Myr)   \\ 
\hline
\begin{tabular}[c]{@{}c@{}}NEOs \\(Atira, Aten, Apollo, Amor)\end{tabular} & Run
  A   & 3020 & 0.43 & 0.0006670  & 1.12    & 0.031    & 23.74    & 0.572 & 12.43  \\
Aten  & Run A & 1000 & 0.25 & 0.0000814  & 8.66 & 0.315  & 24.40 & 0.513  & 16.53 \\
Apollo   & Run A  & 1000 & 0.45 & 0.0005800     & 1.59      & 0.043      & 22.52     & 0.553   & 12.06 \\Amor  & Run A   & 1000 & 0.48 & 0.0011700   & 0.76      & 0.020      & 28.17     & 0.711 & 14.47 \\
\begin{tabular}[c]{@{}c@{}}Atira \\(obliquity = 0 deg.)\end{tabular} & Run B                                & 2000 & 0.04 & 0.0000349  & 4.22      & 0.472      & 24.78     & 0.565  & 14.50  \\
\begin{tabular}[c]{@{}c@{}}Atira \\(obliquity = 90 deg.)\end{tabular} & Run B                                & 2000 & 0.18 &  0.0000429  & 8.15      & 0.430      & 28.26     & 0.698  & 18.21 \\
\begin{tabular}[c]{@{}c@{}}Atira \\(obliquity = 180 deg.)\end{tabular} & Run B                               & 2000 & 0.07 & 0.0000808  & 3.61      & 0.305      & 24.11     & 0.559  & 13.86  \\
\hline
\end{tabular}
    \begin{tablenotes} %Comment section out if you don't want table notes
    \item[a] The fitting parameter in Eq. (\ref{eq2})
    \item[b] The reduced chi-square value calculated from the fitting line
    \item[c] The short-term evolution of half-lifetime 
    \item[d] The uncertainty for short-term evolution of half-lifetime
    \item[e] The long-term evolution of half-lifetime 
    \item[f] The uncertainty for long-term evolution of half-lifetime
    \item[g] The mean of short-term and long-term evolution of half-lifetime
    \end{tablenotes} 
    \end{threeparttable} %Comment section out if you don't want table notes
  \label{table:Tab2.0}
  
  \end{table*}
% \end{sidewaystable}%Comment out if you don't want rotated tables!

% table 3.0 : finding the best fitting of NEOs half-lifetime

\begin{table}
\centering
\label{table:Tab3.0} 
\caption{Using the reduced chi-square statistic to find the best fitting for estimating the short-term and long-term half-lifetime of 3020 real NEOs in the Run A simulation.}
\begin{threeparttable} %Comment section out if you don't want table notes
\begin{tabular}{cccc}
\hline \textit{$\alpha$} $^a$ & \textit{$\chi_\nu $} $^b$   & \textit{$T_1$} $^c$  ~(Myr) & \textit{$T_2$} $^d$  ~(Myr)  \\ 
\hline
0.40 & 0.000790 & 0.89        & 21.99         \\
0.41 & 0.000725 & 0.97        & 22.54         \\
0.42 & 0.000685 & 1.04        & 23.12         \\
0.43 & 0.000667 & 1.12        & 23.74         \\
0.44 & 0.000670 & 1.20        & 24.38         \\
0.45 & 0.000692 & 1.28        & 25.06         \\
0.46 & 0.000731 & 1.37        & 25.78         \\
0.47 & 0.000785 & 1.45        & 26.55         \\
0.48 & 0.000852 & 1.54        & 27.36         \\
0.49 & 0.000932 & 1.63        & 28.22         \\
\hline
\end{tabular}
   \begin{tablenotes} %Comment section out if you don't want table notes
    \item[a] The fitting parameter in Eq. (\ref{eq2})
    \item[b] The reduced chi-square value calculated from the fitting line
    \item[c] The short-term evolution of half-lifetime 
    \item[d] The long-term evolution of half-lifetime 
    \end{tablenotes}
 \end{threeparttable} %Comment section out if you don't want table notes

\end{table}

\subsection{Transfer probability of NEOs and Atiras}
\label{section3.2}
\begin{figure} 
    \centering 
    \includegraphics[width=1.0\columnwidth]{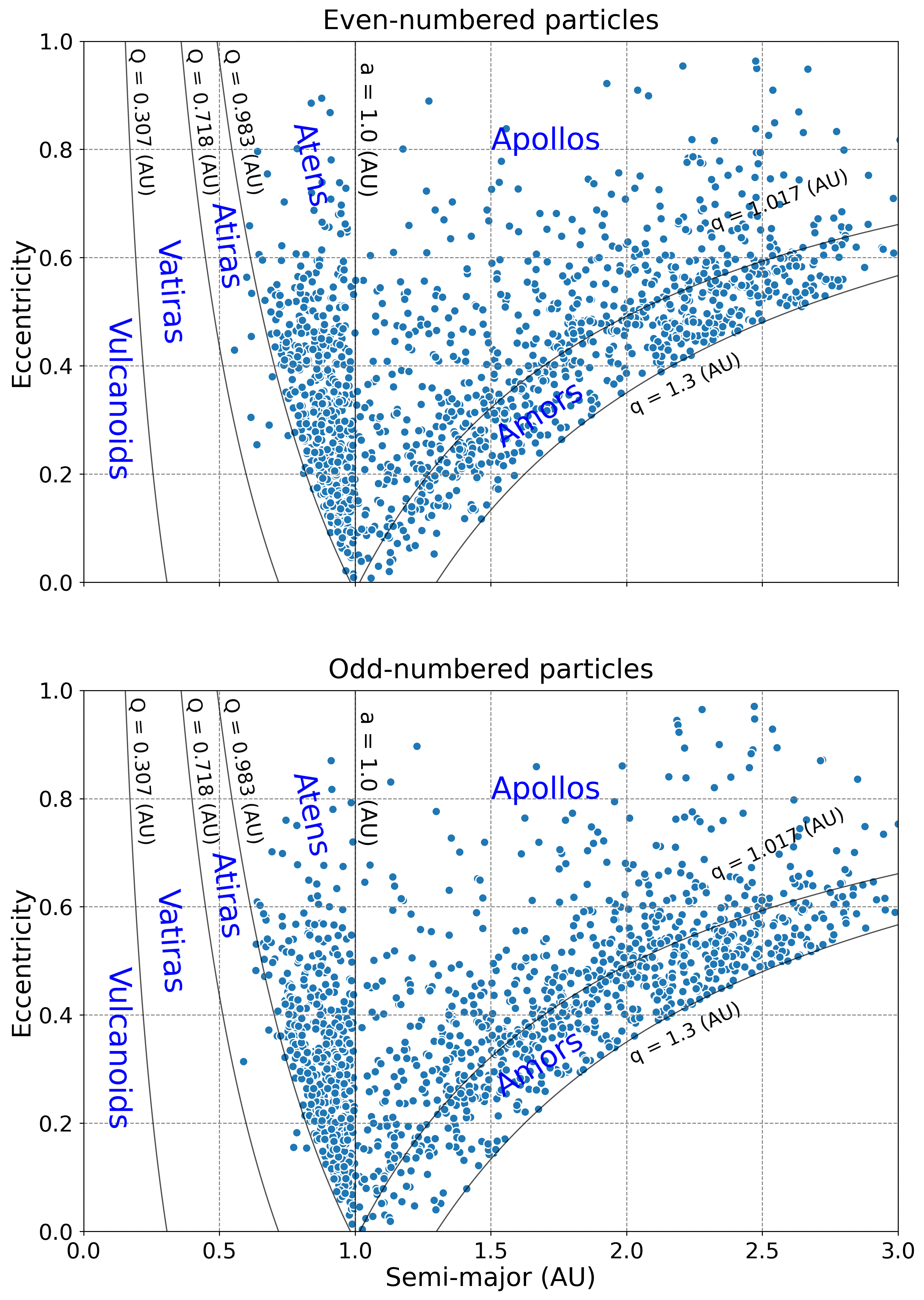}
    \caption{The difference between the odd-numbered particles and the even-numbered particles of the initial conditions in the Run A simulation, which includes Atiras, Atens, Apollos, and Amors. The upper panel is for the orbital distribution of the initial condition of the even-numbered particles while the bottom panel is for the orbital distribution of the initial condition of the odd-numbered particles. }
    \label{fig5}
\end{figure}

The simulation results from \citet{Marcos2020} and \citet{Greenstreet_2020}, both indicated that 2020 $AV_2$ was an Atira-class asteroid before entering the orbital region of the Vatira-class asteroid. Therefore, in order to investigate the relationship between the Vatira-class and Atira-class asteroids, we calculated the transfer probability of each sub-groups of NEOs at the first transition. We tracked each particle in our simulation and measured the fraction of NEOs when it was the first time to evolve from an original class to another class. Taking the Aten-class asteroids as an example (see Table \ref{table:Tab2.0}), we found that $\sim$39.1\% and $\sim$56.2\% of them migrated into the Atira- and Apollo-class asteroids for the first transition, respectively, since it is easier for them to migrate into the two subgroups instead of the other subgroups (i.e. Vatira- and Amor-class asteroids region). It is noticed that $\sim$0.3\% and $\sim$0.5\% of the Aten-class asteroids were ejected into Vatira- and Amor-class asteroids region. Also, about 3.9\% of them collided with planets before entering another class. Furthermore, we split the test particles that were sorted by the date of first observation into even- and odd-numbered particles, allowing the even and odd-numbered groups to give comparable orbital elements with similar initial conditions shown in Figure \ref{fig5}. The method is similar to \citet{greenstreet2012orbital}.
Accordingly, the transfer probability is calculated from the average of the percentage values in the specific NEOs class from two populations of test particles with the probable error of being given by half of the difference between the two respective transfer probabilities.
For example, the transfer probability for asteroids from Aten-class to Apollo-class in the odd-numbered population is given by 55.20 \%, while the one in the even-numbered population is given by 57.2 \%. Therefore, the average transfer probability for asteroids from Aten-class to Apollo-class is (57.20+55.20)/2 \% = 56.20 \%. And the error percentage of the transfer probability is given by (57.20-55.20)/2 \% = 1.000 \%. 

Table \textcolor{blue}{4} indicates that the transfer probability of asteroids from the original class to the other classes of near-Earth objects at the first transition. We found most NEOs of the Aten-class and Apollo-class, tend to migrate outward instead of inward due to locating in the Earth-crossing orbit, so both of them have higher probability to hit planet. Also, a part of the Atens asteroids migrated inward to become Atiras asteroids, in other words, the Vatiras asteroids have the probability of coming from the orbital region of the Aten/Atiras asteroids.
From the result in Table \textcolor{blue}{5}, the majority of the Atira-class asteroids migrated outward instead of inward at the first transition, and the transfer probability of asteroids from Atira-class to Vatira-class for different rotation obliquities (i.e. 0, 90, and 180 degree) are $\sim$13.10, $\sim$13.05, and $\sim$13.25 \%, respectively. It suggests that the non-gravitational Yarkovsky force may play some role in asteroid transportation in the long-term evolution of the inner solar system, even though the transfer probabilities for different rotation obliquities have only a slight difference.

% time-weighted distribution 
\begin{figure}
    \centering 
    \includegraphics[width=0.92\columnwidth]{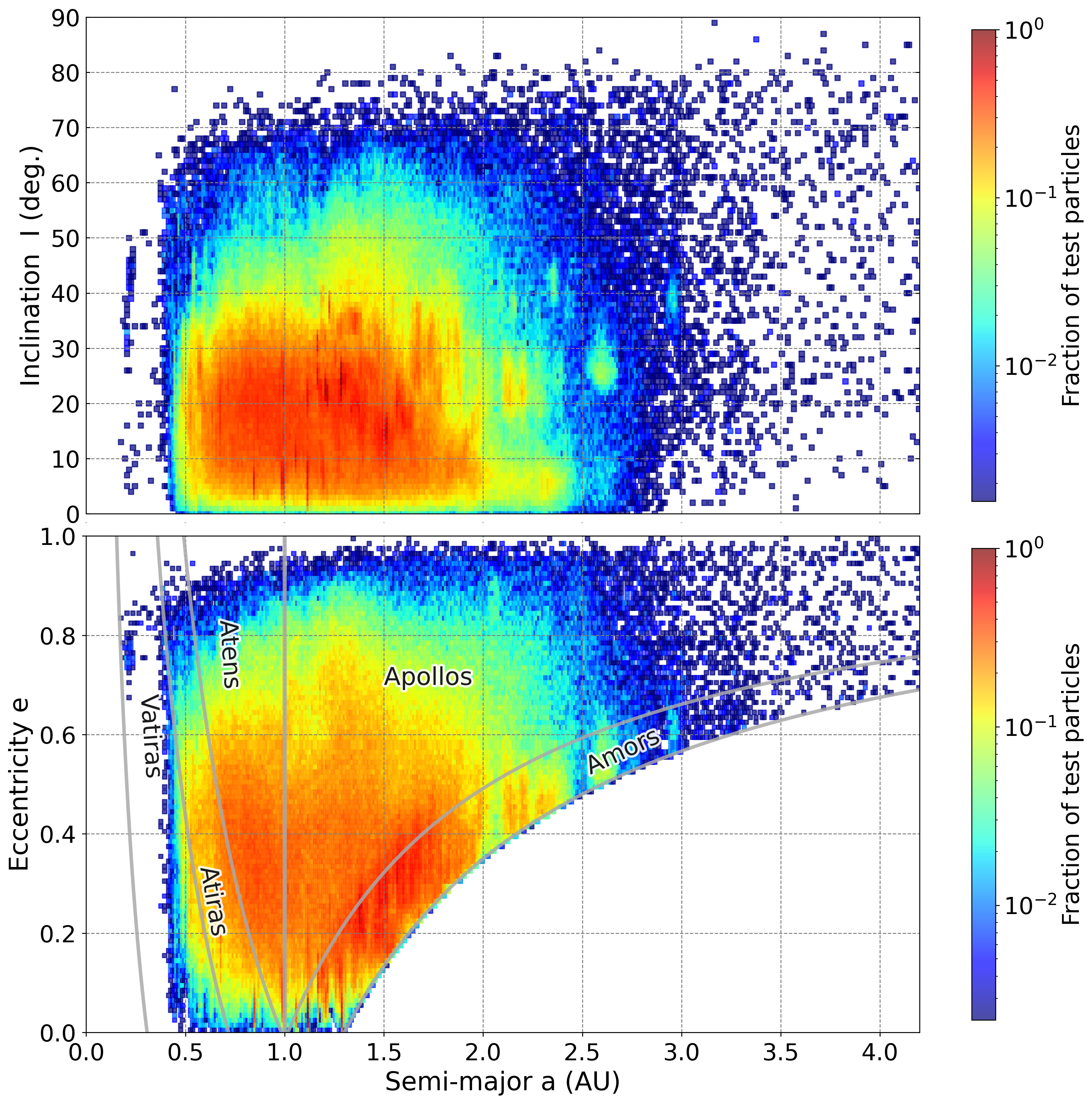}
    \caption{The heat map in the \textit{a-i} space (upper panel) and \textit{a-e} space (lower panel) for recording the orbital evolutionary tracks of the all test particles over the entire orbital evolution. The colour bar represents the fraction of test particles. The bin size for semi-major axis and eccentricity is 0.01, while the bin size of inclination is 1 degree. The solid lines in \textit{a-e} space are used to identify the boundaries of different classes of near-Earth objects and the inner-Venus object: Atira-class (0.718 $<$ \textit{Q} $<$ 0.983 AU), Aten-class (\textit{a} $<$ 1.0 AU, \textit{Q} $>$ 0.983 AU), Apollo-class (\textit{a} $>$ 1.0 AU, \textit{q} $<$ 1.017 AU), Amor-class (1.017 $<$ \textit{q} $<$ 1.3 AU), and Vatira-class (0.307 $<$ \textit{Q} $<$ 0.718 AU). We integrated 3020 samples in the Run A simulation forward in 40 Myr.}
    \label{fig4}
\end{figure}

% table 4.0 : transfer Probability of NEOs (Run A)

\begin{table*}
\centering
\label{table:Tab4.0} 
\caption{The normalized transfer probabilities and errors of asteroids from the original class to the other classes of near-Earth objects at the first transition in the Run A simulation. The values of transfer probabilities are normalized. }
\begin{threeparttable} %Comment section out if you don't want table notes
    \begin{tabular}{ccccccccc} 
    \hline
    \diagbox{Before class}{After class} & Vatira $^a$
      (\%) & Atira $^b$
      (\%) & Aten $^c$
      (\%) & Apollo $^d$ (\%) & Amor $^e$
      (\%) & Hit planet/ Sun $^f$ (\%) & Out of NEO region $^g$ (\%)  \\ 
    \hline
    Aten     & 0.30 $\pm$  0.100 &  39.10 $\pm$  0.500 & 0.00 $\pm$  0.000  & 56.20 $\pm$  1.000  & 0.50 $\pm$  0.100  & 3.90 $\pm$  0.300  & 0.00 $\pm$  0.000    \\
    Apollo   & 0.10 $\pm$  0.100   & 0.80 $\pm$  0.200  & 20.30 $\pm$  0.100 & 0.30 $\pm$  0.300   & 54.50 $\pm$  0.100 & 10.60 $\pm$  0.800 & 13.40 $\pm$  0.400   \\
    Amor     & 0.00 $\pm$  0.000  & 0.00 $\pm$  0.000  & 0.10 $\pm$  0.100  & 58.10 $\pm$  1.300  & 0.70 $\pm$  0.500  & 0.70 $\pm$  0.300  & 40.40 $\pm$  0.600    \\
    \hline
    \end{tabular}
   \begin{tablenotes} %Comment section out if you don't want table notes
    \item[a] Percentage of Vatira-class asteroids with 0.307 AU $<$ \textit{Q} $<$ 0.718 AU.
    \item[b] Percentage of Atira-class asteroids with 0.718 AU $<$ \textit{Q} $<$ 0.983 AU.
    \item[c] Percentage of Aten-class asteroids with \textit{a} $<$ 1.0 AU and \textit{Q} $>$ 0.983 AU.
    \item[d] Percentage of Apollo-class asteroids with \textit{a} $>$ 1.0 AU and \textit{q} $<$ 1.017 AU.
    \item[e] Percentage of Amor-class asteroids with 1.017 AU $<$ \textit{q} $<$ 1.3 AU.
    \item[f] Percentage of test particles is collided with Sun or planets during the first transition.
    \item[g] Percentage of test particles is migrating out of the dynamical region ($q$ $<$ 1.3 AU, $Q$ $>$ 0.718 AU and $a$ $<$ 4.2 AU).
    \end{tablenotes}
 \end{threeparttable} %Comment section out if you don't want table notes
    
\end{table*}

% table 5.0 : transfer Probability of Atira-class asteroids (Run B)

\begin{table*}
\centering
\label{table:Tab5.0} 
\caption{The normalized transfer probabilities and errors of asteroids from the Atira-class to the other classes of near-Earth objects at the first transition in the Run B simulation. The values of transfer probabilities are normalized. }
\begin{threeparttable} %Comment section out if you don't want table notes
    \begin{tabular}{ccccccccc}
    \hline
    \diagbox{Before class}{After class} & Vatira $^a$
      (\%) & Atira $^b$
      (\%) & Aten $^c$
      (\%) & Apollo $^d$ (\%) & Amor $^e$
      (\%) & Hit planet/ Sun $^f$ (\%) & Out of NEO region $^g$ (\%)  \\ 
    \hline
Atira \\  (Obliquity = 0 deg.)     & 13.10 $\pm$ 0.400  & 0.00 $\pm$ 0.000  & 85.00 $\pm$ 0.900 & 0.05 $\pm$ 0.05  & 0.00 $\pm$ 0.000 & 1.85 $\pm$ 0.550    & 0.00 $\pm$ 0.000    \\
Atira\\ (Obliquity = 90 deg.)  & 13.05 $\pm$ 0.005  & 0.00 $\pm$ 0.000  & 84.80 $\pm$ 0.200 & 0.05 $\pm$ 0.05  & 0.00 $\pm$ 0.000 & 2.1 $\pm$ 0.200   & 0.00 $\pm$ 0.000    \\
Atira\\ (Obliquity = 180 deg.)   & 13.25 $\pm$ 0.450  & 0.00 $\pm$ 0.000  & 84.65 $\pm$ 0.450 & 0.00 $\pm$ 0.000  & 0.00 $\pm$ 0.000 & 2.1 $\pm$ 0.000   & 0.00 $\pm$ 0.000   \\ \hline
    \end{tabular}
   \begin{tablenotes} %Comment section out if you don't want table notes
    \item[a] Percentage of Vatira-class asteroids with 0.307 AU $<$ \textit{Q} $<$ 0.718 AU.
    \item[b] Percentage of Atira-class asteroids with 0.718 AU $<$ \textit{Q} $<$ 0.983 AU.
    \item[c] Percentage of Aten-class asteroids with \textit{a} $<$ 1.0 AU and \textit{Q} $>$ 0.983 AU.
    \item[d] Percentage of Apollo-class asteroids with \textit{a} $>$ 1.0 AU and \textit{q} $<$ 1.017 AU.
    \item[e] Percentage of Amor-class asteroids with 1.017 AU $<$ \textit{q} $<$ 1.3 AU.
    \item[f] Percentage of test particles is collided with Sun or planets during the first transition.
    \item[g] Percentage of test particles is migrating out of the dynamical region ($q$ $<$ 1.3 AU, $Q$ $>$ 0.718 AU and $a$ $<$ 4.2 AU).
    \end{tablenotes}
 \end{threeparttable} %Comment section out if you don't want table notes
\end{table*}

\subsection{The number of kilometer-sized inner Venus objects (IVOs) and Atiras from known NEOs population}
Before the discovery of the first Vatira-class asteroid \citep{bolin2020mpec,ip2022discovery}, \citet{granvik2018debiased} and \citet{greenstreet2012orbital} have predicted inner Venus objects to exist according to their NEOs model. Here, we estimated the number of kilometer-sized IVOs and Atiras based on the probability from the Run A simulation and the known NEOs population in JPL's SBDB. Before estimating the potential number of Vatira- and Atira-class asteroids, we set up two assumptions: (1) There is no source region to continuously supply the NEO population, and the test particles in the Run A simulation represent all known NEO populations (see Appendix \ref{Chap:Append_A}). (2) Asteroids with $H$ < 18 are large enough to be well-observed completely. Thereafter, there are 24851 cataloged NEOs with the absolute magnitude of 9.4 $<$ \textit{H} $<$ 33.2 after removing the ineffective samples in the JPL database (see Section \ref{run_A}). Among them, considering the uncertainty of the absolute magnitude $H$ of 2020 $AV_2$ measured by \citet{popescu2020physical}, we found NEOs with 15.625 $<$ \textit{H} $<$ 17.175 occupied $\sim$ 1.7 \% of the known NEOs population. Next, from the heat map of the Run A simulation shown in Figure \ref{fig4}, which recorded the orbital evolutionary tracks of all test particles over the entire orbital evolution of 40 Myr, we estimated the probability of NEOs population to become Vatira-class asteroids to be $\sim$ 0.67 $\pm $ 0.048 \%. As for Atira-class asteroids, the probability from NEOs population is $\sim$ 5.19 $\pm $ 0.085 \%. Moreover, based on the spectroscopic observation, 2020 $AV_2$ is classified as a S-type asteroid \citep{popescu2020physical}. According to the result of \citet{lin2018photometric}, the fractional abundances of the taxonomic complex from 92 NEO samples measured at Lulin Observatory in Taiwan are A-type $\sim$3\%, C-type $\sim$6.5\%, D-type $\sim$8\%, Q-type $\sim$26\%, S-type $\sim$37\%, V-type $\sim$6.5\%, and X-type $\sim$13\%. Finally, we assess that the number of S-type of kilometer-sized would be $\sim$1.05 $\pm$ 0.075 and $\sim$ 8.14 $\pm$ 0.133 for IVOs and Atira-class asteroids, respectively. The result is compatible with 1 known Vatira-class asteroid and 7 known Atira-class asteroids with $H$ < 18 in the JPL database as of September 2022, considering a new Atira-class asteroid 2021 PH27 \footnote[2]{\url{https://minorplanetcenter.net/mpec/K21/K21Q41.html}}  with $H \sim 17.7$ was found on 2021 August 13.
\section{Discussion}
\label{Chap:Chap4}

Although \citet{granvik2018debiased} indicated that the Yarkovsky effect was negligible for the construction of their NEO orbital distribution model on the long-term dynamical evolution of asteroids moving from the main belt into the NEO population. Our result in the second set (Run B) of simulation, namely in Section \ref{section3.2},  suggests the Yarkovsky effects might slightly affect the transportation of the NEO population. Therefore, the Yarkovsky force has a non-negligible effect on Atiras orbital evolution, since inner asteroids with larger da/dt is affected by the rotation obliquity and $a^{-0.5}$ dependence. 

Furthermore, our estimate of the number of kilometer-sized Vatiras and Atiras from the known population is in agreement with the known number of ones in the JPL database as of September 2022. However, the present study focuses solely on the known populations of NEOs. Due to observational biases, the orbital statistic of the Vatiras and Atiras is incomplete. The present survey like the Zwicky Transient Facility twilight survey (ZTF-TS) from \citet{ye2020twilight} indicated they only detected large objects with $H < 20$, $e > 0.2$ and $I >20^{\circ}$ due to Palomar Observatory at the mid-latitude geography location. Also, the expected low-eccentricity stable orbits ($e < 0.2$) identified from the dynamical maps in \citet{ribeiro2016dynamical} and the expected low-inclination Atiras mdoeled in \citet{ye2020twilight} are technically difficult for ZTF-TS to being detected. Despite this discrepancy between ground-based observation and the NEO population model, we can obtain a glimpse of the orbital evolutions of Atiras and other types of NEAs showing how they evolve in general and, in particular, from one dynamical class to another. In future work, we will investigate the Atiras with specific sizes to provide the completeness of the Yarkovsky effect on small asteroids in the inner solar system.

\section{Summary}
\label{Chap:Chap5}
In order to investigate the orbital evolution and relationship of NEOs, forward integrations of Atira-class asteroids for 40 Myr have been carried out by using the \textit{Mercury6} N-body code. The possible non-gravitational effect of the so-called Yarkovsky was also considered. The main results can be summarized as follows:
\begin{enumerate}
  \item We estimated the short-term and long-term half-lifetimes for NEOs and the Atira-class asteroids by the reduced chi-square statistic. The short-term half-lifetime of NEOs is $\sim$ 1.12 Myr, while the long-term half-lifetime is $\sim$23.74 Myr. And the mean half-lifetimes of 12.4 Myr for NEOs is in agreement with $\sim$ 10 Myr median NEO half-lifetime calculated by \citet{gladman1997}. As for the half-lifetime of the Atira-class asteroids, we found that the dynamical half-lifetime of the objects with the Yarkovsky force of $ (da/dt)_0 \sim \pm 9.85 \times 10^{-4} AU Myr^{-1}$  (i.e. the obliquity of 0 and 180 degree) is shorter than those without, since asteroids affected by non-gravitational force tend to migrate into the planet-crossing orbit.
  \item  The transfer probability for the Atira-class transferring to Vatira-class at the first time is about $\sim$13.1 $\pm$ 0.400, $\sim$13.05 $\pm$  0.005, and $\sim$ 13.25 $\pm$ 0.450 $\%$, respectively, depending on the rotation obliquity of 0, 90, 180 deg. It suggests that the radiation force (i.e. $\dot{\Bar{a}} \sim 9.85 \times 10^{-4} AU Myr^{-1}$) may play some role in asteroid transportation in long-term evolution. 
  \item  Based on the probability of NEOs from our simulation and the total number of the known NEOs population cataloged by JPL, our statistical study indicates that there should be 8.14 $\pm$ 0.133 Atira-class asteroids and 1.05 $\pm$ 0.075 Vatira-asteroids of the S-type taxonomy. The values are close to 1 known Vatiras and 7 known Atiras with $H$ < 18 in the JPL database as of September 2022 (see Table \ref{tab:TabA.1}).
  
\end{enumerate}

\section*{Acknowledgements}
This work is supported in part by grant No.107-2119-M-008-012 of the Ministry of Science and Technology, Taiwan. Also, We thank the reviewer for valuable comments and suggestions.

%%%%%%%%%%%%%%%%%%%%%%%%%%%%%%%%%%%%%%%%%%%%%%%%%%
\section*{Data Availability}
The authors confirm that the data supporting the findings of this study are available within the article and its supplementary materials.
 
\bibliographystyle{MNRAS}
\bibliography{mnras_main.bib} % if your bibtex file is called example.bib

%%%%%%%%%%%%%%%%%%%%%%%%%%%%%%%%%%%%%%%%%%%%%%%%%%

%%%%%%%%%%%%%%%%% APPENDICES %%%%%%%%%%%%%%%%%%%%%
% \appendix
% \section{}
\appendix

% ########################  Append_A  #######################
\section{KS test on the orbital and physical property of the two populations}
\label{Chap:Append_A} 
In this section, using the KS test, we compared the JPL population and selected NEO samples from the JPL population whether they are similar. The $p$-value from the KS test is over 0.05, then the selected samples are regarded as being enough to represent all of the known population. Except for the limited number of the Atira-class asteroids (see Section \ref{run_B}), we used the Monte Carlo method to select 1000 samples from the JPL database for other sub-groups of NEO population. In the JPL database, there are 1913, 13785, and 9097 for the Aten-, Apollo- and Amor-class asteroids, respectively. In this statistic, the KS test is applied to the orbital elements of the semi-major axis, eccentricity, inclination, and absolute magnitude. 

\begin{figure}
    \centering 
    \includegraphics[width=1.0\columnwidth]{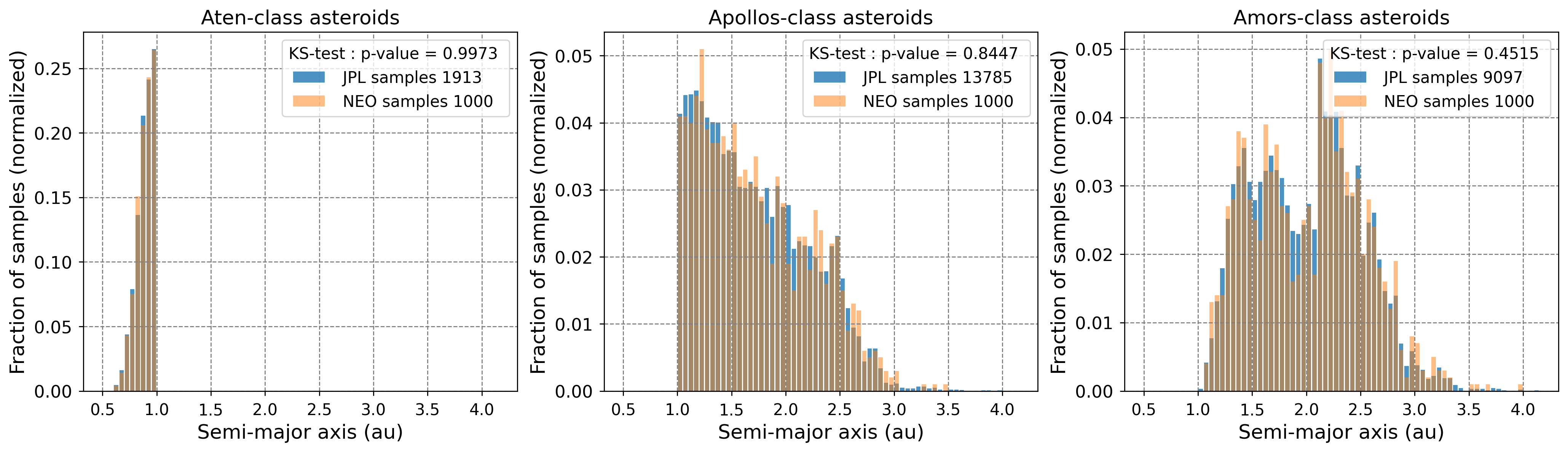}
    \caption{KS test on the semi-major axis of two populations between JPL Database and selected sample from JPL Database. The fraction of the orbital distribution for two populations is normalized. The left-hand panel is for Aten-class asteroids, the middle panel is for Apollo-class asteroids, and the right-hand panel is for Amors-class asteroids. The bin size for the semi-major axis is 0.05 AU. The p-value for the three groups is 1.00, 0.84, and 0.45, respectively, suggesting the similarity of the two populations.}
    \label{figA1}
\end{figure}

\begin{figure}
    \centering 
    \includegraphics[width=1.0\columnwidth]{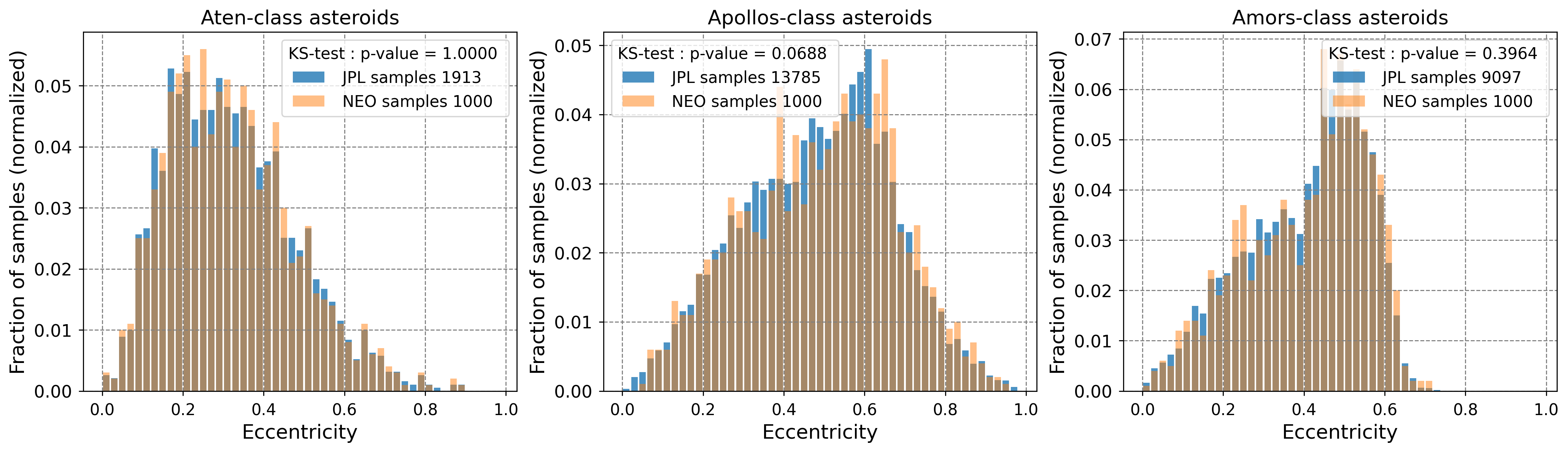}
    \caption{KS test on the eccentricity of two populations between JPL Database and selected sample from JPL Database. The fraction of the orbital distribution for two populations is normalized. The left-hand panel is for Aten-class asteroids, the middle panel is for Apollo-class asteroids, and the right-hand panel is for Amors-class asteroids. The bin size for eccentricity is 0.02. The p-value for the three groups is 1.00, 0.06, and 0.40, respectively, suggesting the similarity of the two populations.}
    \label{figA2}
\end{figure}

\begin{figure}
    \centering 
    \includegraphics[width=1.0\columnwidth]{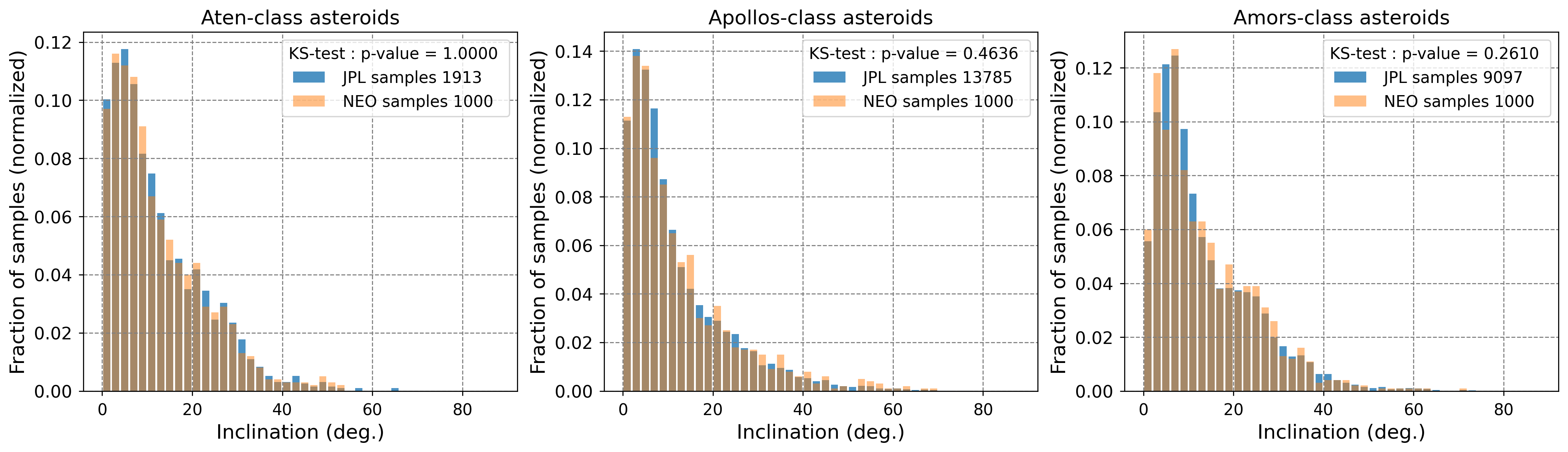}
    \caption{KS test on the inclination of two populations between JPL Database and selected sample from JPL Database. The fraction of the orbital distribution for two populations is normalized. The left-hand panel is for Aten-class asteroids, the middle panel is for Apollo-class asteroids, and the right-hand panel is for Amors-class asteroids. The bin size for inclination axis is 2 degrees. The p-value for the three groups is 1.00, 0.46, and 0.26, respectively, suggesting the similarity of the two populations.}
    \label{figA3}
\end{figure}

\begin{figure}
    \centering 
    \includegraphics[width=1.0\columnwidth]{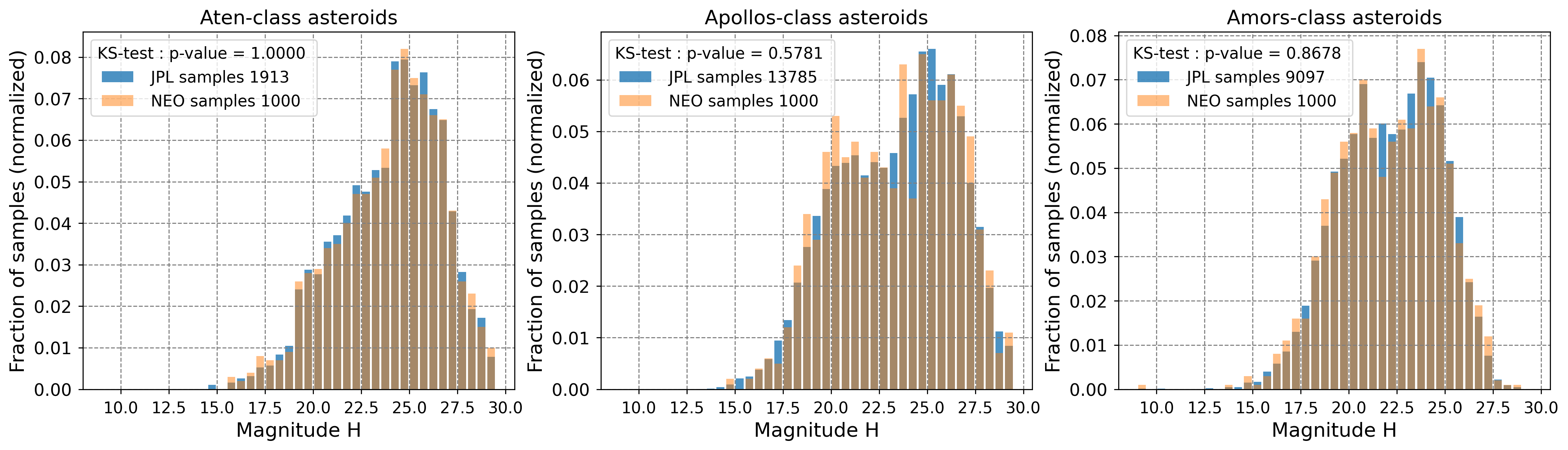}
    \caption{KS test on the absolute magnitude of two populations between JPL Database and selected sample from JPL Database. The fraction of the physical property distribution for two populations is normalized. The left-hand panel is for Aten-class asteroids, the middle panel is for Apollo-class asteroids, and the right-hand panel is for Amors-class asteroids. The bin size for absolute magnitude is 0.5 mag. The p-value for the three groups is 1.00, 0.58, and 0.87, respectively, suggesting the similarity of the two populations.}
    \label{figA4}
\end{figure}

Overall, the $p$-values from the KS test for the Aten-, Apollo- and Amor-class asteroids are higher the criterion we set shown in Figure \ref{figA1}-\ref{figA4}, suggesting the similarity of the two populations and the samples are well-selected to represent the known population. Note that the $p$-value of the eccentricity for Apollos asteroids is lower than other sub-groups since the ratio of the selected samples to the known Apollo-class population in the JPL database is quite low. Similarly, the $p$-values of the four orbital elements of the Atens-class asteroids are almost close to 1.0 due to the higher ratio of the selected samples to the known Aten-class population.  

% fitting result

% fitting result

% fitting result

% fitting result

% ########################  Append_B  #######################
\section{The known Atira-class asteroids identified by JPL on October 30 2020}
\label{Chap:Append_B} 
% \begin{landscape}
% \begin{sidewaystable}[ht!] %Comment out if you don't want rotated tables!
\begin{table*} 
  \centering
  \footnotesize
  \caption{The known Atira-class asteroids identified by JPL on October 30 2020. Noted that asteroid 2020$AV_2$ is classified as the Atira-class asteroid in JPL database, whereas it is classified as the Vatira-class asteroid according to the range of orbital elements ( 0.307 AU $<$ \textit{Q} $<$ 0.718 AU).}
  \begin{threeparttable} %Comment section out if you don't want table notes
  \scalebox{0.95}[1.0]{
  \label{tab:TabA.1}
    \begin{tabular}{llllllllllllc} \hline
    Number & Name             & Spkid  \tnote{a}~ & a  \tnote{c} ~ (AU) & e \tnote{c}~     & $I^{\circ}$  \tnote{c}~  & $\omega^{\circ}$ \tnote{c}~   &  $\Omega^{\circ}$  \tnote{c}  & q  \tnote{c}~ (AU) & Q  \tnote{c}~ (AU) &  $H$  \tnote{c}~  & \# obs. used \tnote{d}~ \\\hline
163693                 & 2003 CP20   & 2163693 & 0.74103 & 0.32212 & 25.61987 & 103.8863 & 252.9120 & 0.50232 & 0.97973 & 16.30 & 617.0      \\
164294                 & 2004 XZ130 & 2164294 & 0.61750 & 0.45461 & 2.94704  & 211.2604 & 5.3191   & 0.33678 & 0.89822 & 20.40 & 72.0          \\
413563                 & 2005 TG45  & 2413563 & 0.68141 & 0.37221 & 23.33621 & 273.4356 & 230.4193 & 0.42778 & 0.93503 & 17.60 & 181.0       \\
418265                 & 2008 EA32  & 2418265 & 0.61593 & 0.30500 & 28.26498 & 100.9563 & 181.8455 & 0.42807 & 0.80380 & 16.40 & 106.0         \\
434326                 & 2004 JG6   & 2434326 & 0.63524 & 0.53110 & 18.94306 & 37.0278  & 352.9976 & 0.29786 & 0.97261 & 18.40 & 138.0         \\
481817                 & 2008 UL90  & 2481817 & 0.69504 & 0.37987 & 24.30966 & 81.1415  & 183.6473 & 0.43102 & 0.95907 & 18.60 & 279.0       \\
& 1998 DK36   & 3184472 & 0.69226 & 0.41602 & 2.01752  & 151.4617 & 180.0427 & 0.40427 & 0.98025 & 25.00 & 4.0                   \\
& ~2006 WE4   & 3360486 & 0.78474 & 0.18292 & 24.76727 & 311.0109 & 318.6144 & 0.64120 & 0.92829 & 18.90 & 59.0                  \\
& 2010 XB11   & 3553148 & 0.61801 & 0.53390 & 29.88699 & 96.3139  & 202.4837 & 0.28806 & 0.94797 & 19.90 & 40.0                  \\
& 2012 VE46   & 3617387 & 0.71303 & 0.36128 & 6.67500  & 8.7797   & 190.5204 & 0.45543 & 0.97064 & 20.20 & 77.0                  \\
& 2013 JX28   & 3638505 & 0.60089 & 0.56404 & 10.76158 & 39.9510  & 354.8904 & 0.26196 & 0.93981 & 20.10 & 55.0                  \\
& 2013 TQ5    & 3648862 & 0.77369 & 0.15557 & 16.39904 & 286.7688 & 247.2960 & 0.65333 & 0.89406 & 19.80 & 47.0                  \\
& 2014 FO47   & 3666783 & 0.75213 & 0.27113 & 19.19713 & 358.6501 & 347.4571 & 0.54821 & 0.95605 & 20.30 & 40.0                  \\
& 2015 DR215  & 3712675 & 0.66641 & 0.47150 & 4.08819  & 314.9523 & 42.2792  & 0.35220 & 0.98063 & 20.30 & 44.0                  \\
& 2015 ME131  & 3791243 & 0.80612 & 0.20422 & 30.23674 & 315.3519 & 162.2739 & 0.64149 & 0.97075 & 19.50 & 6.0                   \\
& 2017 XA1    & 3792468 & 0.80960 & 0.20154 & 17.17614 & 239.6571 & 327.6155 & 0.64644 & 0.97277 & 21.20 & 30.0                  \\
& 2017 YH     & 3824107 & 0.63436 & 0.48255 & 19.83800 & 134.2144 & 147.4802 & 0.32825 & 0.94047 & 18.40 & 72.0                  \\
& 2018 JB3    & 3837637 & 0.68321 & 0.29048 & 40.39004 & 106.4238 & 355.2479 & 0.48475 & 0.88166 & 17.60 & 75.0                  \\
& 2019 AQ3    & 3842903 & 0.58865 & 0.31431 & 47.21887 & 64.4862  & 163.1605 & 0.40363 & 0.77367 & 17.60 & 92.0                  \\
& 2019 LF6   & 3985571 & 0.55542 & 0.42928 & 29.50646 & 179.0286 & 213.7793 & 0.31699 & 0.79385 & 17.20 & 50.0                  \\
& 2020 AV2    & 54016935& 0.55542 & 0.17708 & 15.87226 & 6.7075   & 187.3146 & 0.45706 & 0.65377 & 16.40 & 151.0                 \\
& 2020 HA10   & 54048903& 0.82040 & 0.15444 & 49.66151 & 103.4625 & 27.0801  & 0.69370 & 0.94711 & 19.11 & 23.0                  \\
& 2020 OV1    & 54106533& 0.63751 & 0.25429 & 32.57869 & 296.0217 & 189.8206 & 0.47540 & 0.79963 & 18.68 & 41.0   \\ \hline
    \end{tabular}}
        
    \begin{tablenotes} %Comment section out if you don't want table notes
    \item [a] Object primary SPK-ID
    \item [b] Orbit classification
    \item [c] Orbital elements ($a, e, I, \omega, \Omega, q, Q $) and absolute magnitude ($H$) data from JPL small-body database (\url{https://ssd.jpl.nasa.gov/sbdb.cgi}).
    \item [d] Number of observations (all types) used in fit
    
    \end{tablenotes} 
    \end{threeparttable} %Comment section out if you don't want table notes
\end{table*}%
% \end{sidewaystable}%Comment out if you don't want rotated tables!
% \end{landscape}

%%%%%%%%%%%%%%%%%%%%%%%%%%%%%%%%%%%%%%%%%%%%%%%%%%
% Don't change these lines
\bsp	% typesetting comment
\label{lastpage}
\end{document}